\documentclass[11pt]{article}

\usepackage[T1]{fontenc}
\usepackage[utf8]{inputenc}
\usepackage[a4paper,margin=1in]{geometry}
\usepackage{microtype}
\usepackage{newtxtext,newtxmath}

\usepackage{array}
\usepackage{booktabs}
\usepackage{listings}
\usepackage{enumerate}
\usepackage{graphicx}
\usepackage[font=small,labelfont=bf]{caption}
\usepackage{subcaption}
\usepackage{arydshln}
\usepackage{multirow}
\usepackage{tabularx}
\usepackage{soul}
\usepackage[table]{xcolor}
\usepackage{makecell}
\usepackage[inline]{enumitem}
\usepackage[numbers,sort&compress]{natbib}
\usepackage{xurl}
\usepackage[hidelinks]{hyperref}

\hypersetup{
    pdftitle={Beyond Seeing Is Believing: On Crowdsourced Detection of Audiovisual Deepfakes},
    pdfauthor={Michael Soprano, Andrea Cioci, Stefano Mizzaro}
}

\newcommand{\kr}[1]{}

\newcommand{\ms}[1]{}

\newcommand{\sm}[1]{}

\newcommand{\ac}[1]{}

\newcolumntype{Y}{>{\raggedright\arraybackslash}X}

\definecolor{highlightrow}{HTML}{D9D9D9}

\title{Beyond Seeing Is Believing: On Crowdsourced Detection of Audiovisual Deepfakes}

\author{%
\begin{minipage}{0.95\textwidth}
\centering
\textbf{Michael Soprano}
\quad
\textbf{Andrea Cioci}
\quad
\textbf{Stefano Mizzaro}
\\[0.45em]
\small Department of Mathematics, Computer Science and Physics\\
\small University of Udine, Udine, Friuli-Venezia Giulia, Italy
\\[0.45em]
\small
\href{mailto:michael.soprano@uniud.it}{michael.soprano@uniud.it}
\quad
\href{mailto:cioci.andrea@spes.uniud.it}{cioci.andrea@spes.uniud.it}
\quad
\href{mailto:stefano.mizzaro@uniud.it}{stefano.mizzaro@uniud.it}
\\[0.3em]
\scriptsize
ORCID:
\href{https://orcid.org/0000-0002-7337-7592}{0000-0002-7337-7592}
\quad
\href{https://orcid.org/0009-0005-2539-4813}{0009-0005-2539-4813}
\quad
\href{https://orcid.org/0000-0002-2852-168X}{0000-0002-2852-168X}
\end{minipage}
}

\date{}

\begin{document}

\maketitle

\begin{abstract}
Deepfakes are increasingly realistic and easy to produce, raising concerns about the reliability of human judgments in misinformation settings. We study audiovisual deepfake detection by measuring how consistently crowd workers distinguish authentic from manipulated videos and, when they flag a video as manipulated, how accurately they identify the manipulation type (audio-only, video-only, or audio-video) and how consistently they report manipulation timestamps. We run two matched crowdsourcing studies on Prolific using AV-Deepfake1M and the Trusted Media Challenge (TMC) dataset. We sample 48 videos per dataset (96 total) and collect 960 judgments (10 per video). Results show that crowd workers rarely misclassify authentic videos as manipulated, but they miss many manipulations, and agreement remains limited across videos. Aggregating multiple judgments per video stabilizes the authenticity signal, but it cannot recover manipulations that most workers consistently miss. Manipulation type identification is substantially noisier than authenticity detection even when workers detect a manipulation, with joint audio-video cases being particularly hard to recognize. Overall, these findings suggest that crowdsourcing can provide a scalable screening signal for audiovisual authenticity, while reliable modality attribution remains an open challenge.
\end{abstract}

\vspace{0.5em}
\vspace{0.75em}

\section{Introduction}

Synthetic audiovisual media generated with deep learning techniques, often referred to as deepfakes~\cite{Rana2022DeepfakeDetection}, have rapidly become more realistic and accessible. Consumer-facing tools now enable prompt-based creation and editing of videos and images, for example through text-to-video systems (e.g., Sora)~\cite{openai2024sora} and high-fidelity image generation and editing models (e.g., Gemini)~\cite{raisinghani2025nanobananapro}. By making convincing forgeries easier to produce, these tools amplify misinformation risks and can erode trust in audiovisual reporting. This contributes to a broader ``crisis of knowing'' in which synthetic media undermine the mechanisms through which societies construct shared knowledge, as highlighted by UNESCO~\cite{naffi2025deepfakes}. Recent policy initiatives are putting increasing time pressure on platforms to detect and clearly label such content, highlighting the practical need for scalable authenticity assessment methods~\cite{weatherbed2026deadline}.
 
Threats to the integrity of audiovisual reporting are not limited to AI-generated media. Trust can be damaged even by conventional editing choices, such as splicing fragments of a public speech into an apparently continuous quotation, which may change the perceived meaning and trigger misinformation claims~\cite{bbc_trump_documentary_2025}. More generally, for audiovisual content, the reliability of authenticity judgments depends not only on the available evidence but also on how that evidence is presented and interpreted, motivating a closer look at the human component of detection.

Deepfake detection has largely been addressed as a classification task: models learn to separate authentic from manipulated content by leveraging spatial artifacts, temporal inconsistencies, or audio-video mismatches~\cite{Verdoliva2020MediaForensics,Tolosana2020DeepfakesSurvey}. Deployment-focused work emphasizes detector reliability, with transferability, interpretability, and robustness as key challenges~\cite{WangLiaoChowLinWang2025}. Although detectors can perform well on benchmarks, they often degrade under domain shift, with sharp cross-dataset drops and limited robustness to unseen manipulation techniques~\cite{Nadimpalli2022CrossDataset}. At the same time, studies on human performance and human-machine combinations show that neither humans nor detectors are fully reliable in isolation, and that machine-informed crowds can outperform either component alone~\cite{groh2022deepfake,Diel2024HumanDeepfakeMetaAnalysis}. Consistently, \citet{cooke2025coin} show that unaided human perception can be close to chance under online-like conditions across images, audio, video, and audiovisual media. Human judgments can also be skewed by cognitive biases such as illusory truth from repeated exposure, which is especially relevant for viral deepfakes~\cite{soprano2024cognitivebiases,ahmed2024illusory}. More broadly, crowdsourced judgments can provide a scalable signal for authenticity assessment in misinformation settings, for example as part of human-in-the-loop workflows~\cite{Demartini2020HumanInTheLoop}.

In this work, we focus on the human component of deepfake detection and its role in countering misinformation spread through videos. We run two crowdsourcing tasks to collect, at scale, judgments of authenticity and manipulation type for audiovisual deepfakes. Using a single experimental protocol, we study two benchmarks: AV-Deepfake1M \cite{cai2024av} and the Trusted Media Challenge (TMC) dataset \cite{chen2022trustedmedia}. We analyze workers' accuracy, agreement, and error profiles across manipulation types and datasets to assess whether crowd judgments provide a useful authenticity signal in broader detection settings. Within this framework, our study is guided by the following Research Questions (RQs):
\begin{enumerate}[label={RQ\arabic*}, leftmargin=0.95cm]
\item \label{rq-1} To what extent can a crowdsourcing-based approach correctly distinguish between authentic and manipulated videos?
\item \label{rq-2} How much do workers agree with one another when judging whether a video is authentic or manipulated?
\item \label{rq-3} How accurately can workers identify the manipulation type (audio-only, video-only, or audio-video) for manipulated videos? How consistently do they localize it via timestamps?
\end{enumerate}

Our contributions are threefold: (i) we run two matched Prolific studies on AV-Deepfake1M and TMC and release the resulting authenticity, manipulation type, and timestamp judgments together with the full task configuration, (ii) we analyze accuracy, agreement, and error profiles across datasets and manipulation types under the same protocol and assess the internal consistency of reported manipulation timestamps, and (iii) we compare majority vote and Dempster-Shafer aggregation for authenticity detection and quantify the resulting trade-off between missed manipulations and false alarms.
We release all data and the full task configuration on OSF: \url{https://doi.org/10.17605/OSF.IO/9RJ28}.

Our results show that crowdsourcing provides an authenticity signal for audiovisual deepfakes, but performance depends strongly on the dataset and on the task design. Across both benchmarks, most errors come from missed manipulations rather than false positives, and the difference between AV-Deepfake1M and TMC is large in both accuracy and agreement. Aggregation tends to increase sensitivity to manipulations, but it can also introduce additional false positives, highlighting a practical trade-off when crowd judgments are used as a screening signal. Finally, even when workers detect that a video is manipulated, attributing the anomaly to audio, video, or both remains unreliable, especially for joint audio-video cases; however, when workers do flag a manipulation, their timestamp reports can still converge on a plausible segment.

The remainder of this paper is organized as follows: Section~\ref{sec:related-work} reviews background and related work, Section~\ref{sec:methodology} describes datasets and task design, and Section~\ref{sec:results} presents the results. Section~\ref{sec:discussion} discusses the findings, Section~\ref{sec:limitations} outlines limitations, Section~\ref{sec:implications} summarizes practical implications, and Section~\ref{sec:conclusions-future} concludes with directions for future work. This work was prepared for the ROMCIR 2026 workshop~\cite{pichel2026romcir}.

\section{Background and Related Work}
\label{sec:related-work}

We first review crowdsourcing-based authenticity and truthfulness judgments in misinformation settings (Section~\ref{sec:related-work-subsec:crowdsourcing-truthfulness}), then discuss human factors in deepfake detection (Section~\ref{sec:related-work-subsec:deepfakes-human-factors}), and finally summarize the audiovisual deepfake datasets and benchmarks most relevant to our study (Section~\ref{sec:related-work-subsec:deepfake-datasets}).

\subsection{Crowdsourcing Authenticity and Truthfulness}
\label{sec:related-work-subsec:crowdsourcing-truthfulness}

Crowdsourcing plays a central role in tasks that require subjective judgment in misinformation settings, including authenticity and truthfulness assessments. \citet{he2025crowdsurvey} survey the roles that crowds can play in combating online misinformation across task designs and pipelines.

Prior work shows that aggregated non-expert judgments can serve as a proxy for professional fact-checking and support scalable screening pipelines. For example, crowd assessments of news source quality correlate with expert assessments, and small crowds can match fact-checkers when judging the accuracy of news content~\cite{pennycook2019fighting,allen2021scaling}. Consistent with these findings, \citet{roitero2020crowd} and \citet{10.1145/3546917} show that crowd labels, when properly collected and aggregated, align with expert verdicts and can be used to train or evaluate automated fact-checking systems. Beyond binary authenticity, research has also considered richer notions: \citeauthor{SOPRANO2021102710} \cite{SOPRANO2021102710, soprano2025crowdveritas} introduce a multidimensional perspective and show that assessors can judge several facets of a claim (e.g., correctness, completeness, and trustworthiness), while \citet{10.1145/3531146.3534629} analyze how individual biases affect judgment quality and the resulting aggregated labels. More recently, \citet{BARBERA2024103792} synthesize evidence from a series of crowdsourcing-based fact-checking experiments and conclude that, under appropriate task design and quality-control mechanisms, crowdsourcing is a viable methodology to tackle misinformation at scale.

A complementary line of research enriches authenticity and truthfulness assessments with additional signals. For example, \citeauthor{roitero2021crowd} \cite{roitero2021crowd, 10.1145/3340531.3412048} collect self-reported confidence, background information, and behavioral logs (e.g., queries, clicked URLs, and justifications) to study evidence-seeking behavior and its relationship with judgment quality. \citet{doi:10.1177/17456916231190388} show that aggregated crowd judgments can identify low-quality and misleading news sources at scale, while \citet{roitero2025collecting} and \citet{10.1145/3726302.3730091} explore richer assessment approaches, such as collecting multiple complementary signals and using magnitude estimation scales instead of coarse ordinal labels. More recently, hybrid pipelines have explored combining large language models with crowdsourcing signals for misinformation detection~\cite{zeng2024llmcrowd}.

Similar methodologies have also been adopted in broader information quality settings beyond fact-checking. \citeauthor{CEOLIN2022102107} \cite{CEOLIN2022102107, 10.1007/978-3-030-74296-6_6} ask assessors to evaluate review texts along multiple quality dimensions (e.g., accuracy, completeness, and credibility) and combine these judgments with formal argumentation models to obtain aggregate assessments that correlate with user signals such as helpfulness votes. Relatedly, alternative aggregation schemes have been studied in crowdsourcing-based peer review approaches~\cite{soprano2025readersourcing}.

Beyond textual settings, crowdsourcing-based approaches have also been used to collect human judgments on deepfakes and to study how human-side factors relate to authenticity decisions~\cite{salini2024crowdcomputingdeepfake}. Relatedly, large-scale crowd annotations have been collected to model human perception of deepfake realism, for example through benchmarks based on realism scores and artifact descriptions~\cite{peng2025dream}.

Building on this line of work, we apply crowdsourcing to audiovisual deepfake assessment and study the reliability of authenticity and manipulation type judgments.

\subsection{Human Factors in Deepfake Detection}
\label{sec:related-work-subsec:deepfakes-human-factors}

Human performance in deepfake detection is often limited, and accuracy can remain close to chance under online-like conditions across modalities~\cite{cooke2025coin,Diel2024HumanDeepfakeMetaAnalysis}. The meta-analysis by \citet{Diel2024HumanDeepfakeMetaAnalysis} shows that manipulated videos are harder to identify than authentic videos and that brief training or feedback leads to only modest gains. Overall, performance varies substantially across videos and usage contexts.

Experimental work provides more evidence on the factors that shape judgments and on the potential for intervention. \citet{korshunov2020deepfake} compare human observers and automatic detectors and show that high-quality manipulations easily fool most viewers, with poor calibration between confidence and correctness. \citet{kobis2021fooled} study audiovisual political deepfakes and find that participants detect manipulations only slightly above chance even when warnings or incentives are provided, again pointing to miscalibration. In the political domain, \citet{appel2022political} propose a theoretical model of deepfake detection and report three preregistered studies on political deepfake videos. Similarly, \citet{somoray2023strategies} evaluate strategy-based interventions and report, at best, small improvements, indicating that simple instructions are insufficient to substantially improve detection.

Beyond judgments made without automated support, several studies highlight complementary error profiles between humans and detectors, motivating hybrid human-AI approaches. \citet{groh2022deepfake} show that combining human and model predictions can improve performance, although misleading model outputs can also push users towards incorrect decisions. In audiovisual settings, \citet{groh2024politicalspeech} report that presenting both channels can support better identification than text alone, while highly realistic synthetic audio remains particularly challenging. Together, these results indicate that modality matters, but audiovisual presentation alone does not eliminate systematic errors.

Compared to this body of work, we study crowd judgments under a matched audiovisual protocol on two benchmarks and focus on both authenticity detection and manipulation type attribution. We quantify accuracy, agreement, and error profiles to assess whether crowdsourcing provides a useful screening signal for audiovisual authenticity.

\subsection{Deepfake Datasets and Benchmarks}
\label{sec:related-work-subsec:deepfake-datasets}

\begin{table}[t]
\centering
\small
\caption{Key deepfake datasets and benchmarks. Reported fields include year, modality, user-study reporting, collection setting, and scale. Bold rows highlight the two datasets used in this study.}
\label{tab:deepfake-datasets-summary}

{\setlength{\tabcolsep}{4pt}
\begin{tabularx}{\linewidth}{Y c c c p{2cm} Y}
\toprule
\textbf{Dataset} & \textbf{Year} & \textbf{Modality} & \textbf{User Study} & \textbf{Setting} & \textbf{Scale} \\
\midrule
DeeperForensics-1.0 \cite{jiang2020deeperforensics} & 2020 & V & Yes & Controlled & 60{,}000 videos \\
DFDC \cite{dolhansky2020dfdc} & 2020 & V & No & Curated & 128{,}154 videos \\
ForgeryNet \cite{he2021forgerynet} & 2021 & I+V & No & Online & 221{,}247 videos \\
FFIW-10K \cite{zhou2021faceforensicswild} & 2021 & V & Yes & Online & 20{,}000 videos \\
FakeAVCeleb \cite{khalid2021fakeavceleb} & 2021 & A+V & No & Online & 20{,}000 videos \\
KoDF \cite{kwon2021kodf} & 2021 & V & Yes & Curated & 237{,}942 videos \\
DF-Mobio \cite{korshunov2022dfmobio} & 2021 & V & No & Online & $\approx$46{,}000 videos \\
LAV-DF \cite{cai2023lavdf} & 2022 & A+V & No & Curated & 136{,}304 videos \\
\textbf{TMC} \cite{chen2022trustedmedia} &
\textbf{2022} &
\textbf{A+V} &
\textbf{Yes} &
\textbf{Curated} &
\textbf{6{,}943 videos} \\
DF-Platter \cite{narayan2023dfplatter} & 2023 & V & Yes & Online & 133{,}260 videos \\
\textbf{AV-Deepfake1M} \cite{cai2024av} &
\textbf{2024} &
\textbf{A+V} &
\textbf{No} &
\textbf{Curated} &
\textbf{1{,}146{,}760 videos} \\
Deepfake-Eval-2024 \cite{chandra2025deepfakeeval} & 2025 & A+V+I & No & Online & 2{,}036 videos \\
\bottomrule
\end{tabularx}}
\end{table}

Deepfake detection is typically evaluated on benchmarks that differ substantially in scale, realism, and modality. Early large-scale datasets such as DeeperForensics-1.0 \cite{jiang2020deeperforensics} and the DeepFake Detection Challenge (DFDC) dataset \cite{dolhansky2020dfdc} focus on visual face forgery in controlled or semi-controlled settings, while introducing perturbations and acquisition variability to encourage robustness. Subsequent benchmarks, including ForgeryNet \cite{he2021forgerynet}, FFIW-10K \cite{zhou2021faceforensicswild}, KoDF \cite{kwon2021kodf}, DF-Mobio \cite{korshunov2022dfmobio}, and DF-Platter \cite{narayan2023dfplatter}, increase scale, introduce more diverse scenes, and motivate evaluations of cross-dataset generalization. At the same time, many benchmarks remain video-centric and often emphasize specific manipulation families, which contributes to sharp performance drops under distribution shift.

A complementary line of work proposes multimodal benchmarks that integrate audio and video manipulations. FakeAVCeleb \cite{khalid2021fakeavceleb} provides an audiovisual deepfake corpus, while LAV-DF \cite{cai2023lavdf} targets short, localized audiovisual forgeries and supports temporal localization tasks. The Trusted Media Challenge (TMC) dataset \cite{chen2022trustedmedia} provides a multimodal benchmark with diverse speakers and realistic perturbations, and has been used to evaluate both human and automatic detectors in a competition setting. More recently, AV-Deepfake1M \cite{cai2024av} introduces a large-scale audiovisual dataset with rich labels, and Deepfake-Eval-2024 \cite{chandra2025deepfakeeval} collects manipulations circulated online in 2024 (released in 2025), highlighting severe performance drops when moving from laboratory benchmarks to open web content. Recent multilingual audio-video benchmarks further broaden this space \cite{hou2024polyglotfake, croitoru2025mavosdd}.

Table~\ref{tab:deepfake-datasets-summary} provides a comparison of these benchmarks. We indicate modalities using the shorthand I (images), V (video), and A (audio); some datasets include multiple channels (e.g., A+V). We also summarize user-study reporting, setting, and scale. The setting distinguishes controlled acquisition protocols, curated datasets assembled from selected sources, and datasets collected online. Guided by this comparison, we focus on AV-Deepfake1M \cite{cai2024av} and TMC \cite{chen2022trustedmedia}, which share explicit authentic/manipulated labels and a consistent manipulation type taxonomy, enabling a matched sampling design. Deepfake-Eval-2024 \cite{chandra2025deepfakeeval} is a more recent alternative, but its documentation does not provide the details needed to construct an equivalent pool.

\section{Methodology}
\label{sec:methodology}

We first describe the two audiovisual deepfake datasets and how we construct two balanced video pools (Section~\ref{sec:methodology-subsec:dataset}). We then detail the two Prolific studies and the task design (Section~\ref{sec:crowdsourcing-task}). Next, we describe how we aggregate worker judgments into video-level labels for the analyses (Section~\ref{sec:aggregation}). We then introduce the evaluation measures used to assess classification performance, worker agreement, and timestamp consistency (Section~\ref{sec:measures}). Finally, we outline the statistical analysis (Section~\ref{sec:methodology-subsec:stats}).

\subsection{Dataset}
\label{sec:methodology-subsec:dataset}

We draw our experimental samples from two multimodal audiovisual deepfake datasets discussed in Section~\ref{sec:related-work-subsec:deepfake-datasets}: AV-Deepfake1M \cite{cai2024av} and the Trusted Media Challenge (TMC) dataset \cite{chen2022trustedmedia}. We rely on the dataset-provided ground truth and do not manually simulate or edit content. We consider four conditions: authentic (unaltered audio and video), audio-only manipulation, video-only manipulation, and joint audio-video manipulation. To ensure feasibility in a crowdsourcing setting, we restrict the candidate pools to videos playable in a browser-based interface and containing both an audio and a video track, and we perform stratified random sampling by condition.

In each dataset, we sample $12$ videos for each of the four conditions (48 per dataset), for a total of 96 videos overall. The sampled clips have a median duration of $7.0\,\mathrm{s}$ (IQR: $6.0$--$9.0\,\mathrm{s}$) in AV-Deepfake1M and $55.9\,\mathrm{s}$ (IQR: $49.9$--$59.2\,\mathrm{s}$) in TMC. These balanced pools support controlled comparisons, but they are not intended to reflect the prevalence of manipulation types in the full datasets. The resulting samples form the basis of the AV-Deepfake1M-based and TMC-based crowdsourcing tasks described in Section~\ref{sec:crowdsourcing-task}.

\subsection{Crowdsourcing Task}
\label{sec:crowdsourcing-task}

We conducted our data collection on Prolific \cite{PALAN201822} in two stages: (i) an AV-Deepfake1M-based pilot study to validate and refine the protocol, and (ii) two final crowdsourcing tasks, one per dataset (AV-Deepfake1M and TMC).\footnote{We referred to the studies as V1 (AV-Deepfake1M) and V2 (TMC). These identifiers may also appear in the released data.} In both tasks, workers received a brief introduction and detailed instructions and then watched a sequence of videos. For each video, they provided an authenticity judgment. An authentic judgment indicates that neither channel appears altered, whereas a manipulated judgment indicates that at least one channel appears altered. In the latter case, workers also selected the manipulation type and marked an approximate timestamp for when the manipulation appeared to occur. Timestamps were collected with centisecond resolution (\textsf{mm:ss.cc}). Overall, the two-step interface defines four possible outcomes at the judgment level: authentic, audio-only, video-only, and audio-video. At the start of each task, workers also completed a short demographic questionnaire. We complement these self-reported answers with Prolific-provided demographic attributes when available and when consent is granted; some fields are missing because of consent revocation or expired data.

Before launching the final tasks, we conducted a pilot with 75 completed work units (8 videos each; 600 judgments) to assess feasibility and refine the protocol. Pilot data were used only to validate the protocol and inform task design. Pilot results suggested that longer assignments increased cognitive load and were associated with more missed manipulations, so we reduced the assignment length to 4 videos per work unit to limit fatigue and provide a more consistent decision context across items. In the final deployment, we kept the interface, instructions, and label space unchanged for comparability, set a target completion time of 10 minutes, and paid 1.50 GBP per work unit (9.00 GBP/hour).

Data collection for the two final tasks took place on October 20, 2025, in two separate windows: 10:44--11:37 (CEST) for the AV-Deepfake1M-based task and 13:19--23:03 (CEST) for the TMC-based task. Overall, we collected 240 completed work units (120 per task), corresponding to 960 video-level judgments (480 per task). Within each task, the 48 sampled videos were distributed across work units so that each video received 10 independent judgments from distinct workers, and the order of videos within each work unit was randomized to reduce learning effects. Participation statistics and worker demographics are reported in Section~\ref{sec:results-subsec:demographics} (Table~\ref{tab:demographics-v1-v2}).

Completion times were skewed toward longer durations. Across both tasks, the mean completion time was 8 minutes 13 seconds ($\sigma$ = 4 minutes 33 seconds) and the median was 6 minutes 39 seconds; given the fixed reward of 1.50 GBP per work unit, this implies a median effective rate of approximately 13.53 GBP/hour. By task, the median time was 5 minutes 56 seconds (IQR: 5:02--8:08) for AV-Deepfake1M and 9 minutes 38 seconds (IQR: 8:54--12:17) for TMC, consistent with the longer videos in TMC.

We designed and conducted all tasks using the Crowd\_Frame framework~\cite{10.1145/3488560.3502182}, which facilitates the deployment of crowdsourcing experiments.

\subsection{Judgment Aggregation}
\label{sec:aggregation}

Each video receives ten judgments (Section~\ref{sec:crowdsourcing-task}), which we aggregate into a single authenticity label for the video-level analyses in Section~\ref{sec:results-subsec:df-accuracy} and Section~\ref{sec:results-subsec:consistency-crowd} using majority vote (Section~\ref{sec:aggregation-majority}) or Dempster-Shafer (Section~\ref{sec:aggregation-ds}). For manipulation type analyses in Section~\ref{sec:results-subsec:manipulation-type}, we retain the four labels induced by the two-step interface.

\subsubsection{Majority Vote}
\label{sec:aggregation-majority}

For each video, we count how many workers assign each authenticity label and select the label with the highest number of votes. When individual judgments are noisy but centered around the true label, aggregation can reduce variance and provide a more stable estimate than any single judgment~\cite{Penrose1946}. Given its simplicity, majority vote is commonly adopted as a baseline aggregation strategy in crowdsourcing studies~\cite{roitero2020crowd, BARBERA2024103792, roitero2021crowd, 10.1007/978-3-642-28997-2_16}.

In the event of a tie (e.g., $5$ vs.\ $5$ with ten votes), the majority label is undefined and a deterministic rule is needed. We break ties in favor of the authentic class, making aggregation conservative with respect to false positives. For manipulation type aggregation, ties are broken deterministically by selecting the smallest label value.

\subsubsection{Dempster-Shafer}
\label{sec:aggregation-ds}

We also compute an alternative aggregated authenticity label using Dempster-Shafer (DS) theory~\cite{dempster1967upper, shafer1976mathematical}. DS aggregation treats each worker's authenticity response as evidence for one of the two outcomes while explicitly representing uncertainty. Concretely, each judgment provides weighted support for the selected label and assigns the remaining weight (or mass, in DS terminology) to uncertainty. Belief-function aggregation has also been explored in crowdsourcing to combine imperfect worker answers and model contributor reliability~\cite{thierry2023monitor, abassi2017gold}.

We set the evidence weight based on worker reliability, computed within each final task using a leave-one-out procedure over the worker's other authenticity answers, and use it to weight the current judgment. To avoid giving extra influence to systematically low-performing workers, we lower-bound the weight at chance so that such judgments contribute only uninformative evidence. For each video, we then combine the evidence from all workers using Dempster's rule of combination~\cite{dempster1967upper}.

To obtain a final decision comparable to majority vote, we convert the aggregated evidence into a probability distribution using the pignistic transformation~\cite{smets1994transferable} and assign the label with the highest resulting probability, breaking ties deterministically as in majority vote.

\subsection{Evaluation Measures}
\label{sec:measures}

We describe the measures used to evaluate the outcomes of our crowdsourcing tasks. We first present the classification measures used for authenticity detection and manipulation type identification (Section~\ref{sec:metrics-subsec:classification}), then introduce the worker agreement measures used to quantify the internal consistency of authenticity judgments (Section~\ref{sec:metrics-subsec:agreement}), and finally define the timestamp consistency measures used to analyze the temporal localization of manipulations (Section~\ref{sec:metrics-subsec:timestamps}).

\subsubsection{Classification Metrics}
\label{sec:metrics-subsec:classification}

We evaluate classification performance using standard binary metrics derived from the confusion matrix. Using the video-level labels obtained via the aggregation strategies described in Section~\ref{sec:aggregation}, we treat manipulated videos as the positive class and authentic videos as the negative class, and compute precision, recall, false positive rate, false negative rate, and F1 score using their standard definitions.

For manipulation type identification, we report video-level accuracy on ground truth manipulated videos under three evaluation settings. In \textsf{Authenticity + Type}, an authentic judgment is treated as an implicit \textsf{Real} prediction, so missed manipulations count as type errors. This is the most stringent setting, as it penalizes both missed manipulations and incorrect type assignments. In \textsf{Any Fake Vote}, we derive the manipulation type using only judgments labeled as manipulated and include a video as long as at least one worker flags it as manipulated. In \textsf{Majority Fake}, we evaluate type accuracy only for videos whose majority authenticity label is manipulated, thus focusing on cases where the crowd detects manipulation after aggregation.

\subsubsection{Worker Agreement}
\label{sec:metrics-subsec:agreement}

We use agreement measures to quantify the internal consistency of the crowdsourced labels, independently of the ground truth. Agreement with the ground truth is captured by the classification metrics reported in Section~\ref{sec:metrics-subsec:classification}. We rely on three complementary measures commonly used in crowdsourcing studies~\cite{roitero2020crowd,SOPRANO2021102710,10.1145/3531146.3534629,BARBERA2024103792,roitero2021crowd,10.1145/3340531.3412048,10.1145/3726302.3730091,10.1007/978-3-030-45442-5_26}: Krippendorff's $\alpha$~\cite{krippendorff2011computing}, majority agreement, and pairwise agreement~\cite{maddalena2017considering}. We compute all three measures from worker judgments only.

Krippendorff's $\alpha$ is a chance-corrected coefficient for inter-annotator agreement. In general, higher values indicate stronger agreement, with $\alpha = 1$ corresponding to perfect agreement. In our study, it provides a global summary of how reliably workers label the videos. Majority agreement is defined as, for each video, the fraction of workers who select the majority label, averaged across videos. Following \citet{maddalena2017considering}, pairwise agreement is the average, across videos, of the proportion of unordered worker pairs that assign the same label; compared to majority agreement, it provides a more fine-grained view of consistency and is less sensitive to class imbalance.

\subsubsection{Timestamp Consistency}
\label{sec:metrics-subsec:timestamps}

For timestamp analyses, we consider the timestamps provided in manipulated judgments and normalize them by the video duration to enable comparisons across short and long clips. 

We quantify timestamp consistency at the video level using the median and interquartile range (IQR) of the normalized timestamps and a timestamp agreement score, defined as the fraction of timestamp votes that fall within a $\pm 5\%$ window of the video duration around the per-video median. We compute these measures on ground truth manipulated videos that receive at least three timestamp annotations from manipulated judgments.

\subsection{Statistical Analysis}
\label{sec:methodology-subsec:stats}

For inferential analyses, we treat each sampled video as one observation and compare the corresponding per-video distributions. Since these measures are not guaranteed to be normally distributed and the sample size is limited (Section~\ref{sec:methodology-subsec:dataset}), we rely on nonparametric tests that do not require normality assumptions and are robust to skewed distributions.

To compare AV-Deepfake1M and TMC, which are independent samples, we use Mann-Whitney U tests~\cite{mann1947test}. To compare two methods on the same set of videos within a dataset, for paired outcomes such as majority vote vs.\ DS on per-video correctness, we use McNemar's test~\cite{mcnemar1947note}. When comparing more than two groups within the same dataset, for example across manipulation types, we use the Kruskal-Wallis test~\cite{kruskal1952use}; when it is significant, we follow up with pairwise Mann-Whitney tests~\cite{mann1947test}.

When multiple hypotheses are tested within the same analysis block, we control the family-wise error rate through multiple-comparison correction of the corresponding $p$-values. We use a Bonferroni correction~\cite{bland1995multiple} for small, predefined families of tests, and Holm-Bonferroni for post-hoc pairwise comparisons following Kruskal-Wallis~\cite{holm1979simple}. We report adjusted $p$-values and summarize significance using conventional threshold notation (e.g., $p<0.05$, $p<0.01$). Alongside medians and significance outcomes, we summarize the magnitude and direction of differences using Cliff's $\delta$~\cite{cliff1993dominance} and compute confidence intervals for $\delta$ using a nonparametric bootstrap over videos~\cite{efron1993bootstrap}.

\section{Results}
\label{sec:results}

We first summarize participation and worker demographics (Section~\ref{sec:results-subsec:demographics}). We then answer \ref{rq-1}--\ref{rq-3} by analyzing authenticity detection accuracy (\ref{rq-1}, Section~\ref{sec:results-subsec:df-accuracy}), the consistency of crowd judgments (\ref{rq-2}, Section~\ref{sec:results-subsec:consistency-crowd}), and manipulation type identification (\ref{rq-3}, Section~\ref{sec:results-subsec:manipulation-type}).

\subsection{Worker Statistics and Demographics}
\label{sec:results-subsec:demographics}

\begin{table}[t]
\centering
\footnotesize
\caption{Workers' demographics for both tasks. Top: self-reported questionnaire. Bottom: Prolific-provided demographics. Abbreviations: \textsf{PNS} = Prefer Not to Say, \textsf{CR} = Consent Revoked, \textsf{Exp.} = Expired.}
\label{tab:demographics-v1-v2}
\setlength{\tabcolsep}{4.2pt}
\renewcommand{\arraystretch}{1.12}
\begin{tabularx}{\linewidth}{@{}>{\raggedright\arraybackslash}p{1.75cm} >{\raggedright\arraybackslash}X >{\raggedright\arraybackslash}X@{}}
\toprule
\textbf{Item} & \textbf{AV-Deepfake1M Task} & \textbf{TMC Task} \\
\midrule
\textsf{Age} &
5.8\% 19--25, 28.3\% 26--35, 43.3\% 36--50, 22.5\% 51--80 &
5.0\% 19--25, 21.7\% 26--35, 38.3\% 36--50, 35.0\% 51--80 \\
\textsf{Education} &
16.7\% High School or less, 26.7\% College/Assoc., 36.7\% Bachelor's, 20.0\% Postgraduate &
18.3\% High School or less, 27.5\% College/Assoc., 38.3\% Bachelor's, 15.8\% Postgraduate \\
\textsf{Party} &
30.0\% Rep., 33.3\% Dem., 35.8\% Indep., 0.8\% Other &
32.5\% Rep., 40.0\% Dem., 25.8\% Indep., 1.7\% Other \\
\textsf{Ideology} &
9.2\% Very Cons., 25.0\% Conservative, 23.3\% Moderate, 25.8\% Liberal, 16.7\% Very Lib. &
10.0\% Very Cons., 22.5\% Conservative, 21.7\% Moderate, 32.5\% Liberal, 12.5\% Very Lib., 0.8\% n/a \\
\textsf{Income} &
15.8\% $\le$ \$30k, 45.0\% \$30k--\$75k, 39.1\% \$75k or more &
18.3\% $\le$ \$30k, 37.5\% \$30k--\$75k, 44.2\% \$75k or more \\
\addlinespace[3pt]
\cdashline{1-3}
\addlinespace[3pt]
\textsf{Gender} &
52.1\% Female, 47.9\% Male &
49.2\% Female, 49.2\% Male, 0.8\% \textsf{PNS}, 0.8\% \textsf{CR} \\
\textsf{Ethnicity} &
75.6\% White, 12.6\% Asian, 8.4\% Black, 0.8\% Mixed, 2.5\% Other &
67.5\% White, 12.5\% Black, 7.5\% Asian, 6.7\% Mixed, 4.2\% Other, 0.8\% \textsf{Exp.}, 0.8\% \textsf{CR} \\
\textsf{Residence} &
100\% US &
99.2\% US, 0.8\% \textsf{CR} \\
\textsf{Birth} &
90.8\% US; 1.7\% CN; 1.7\% IN; Others 0.8\% &
83.3\% US; 2.5\% CA; 1.7\% UK; 1.7\% CN; Others 1.6\% \\
\textsf{Nationality} &
89.9\% US; 1.7\% IN; Others 0.8\%; 0.8\% \textsf{Exp.} &
93.3\% US; 1.7\% UK; Others 0.8\%; 0.8\% \textsf{CR} \\
\textsf{Language} &
89.1\% EN; 4.2\% ES; 1.7\% ZH; Others 0.8\% &
92.5\% EN; 1.7\% ES; 1.7\% AR; 1.7\% ZH; Others 1.6\% \\
\textsf{Fluency} &
81.5\% EN only; 18.5\% EN+ &
85.8\% EN only; 13.3\% EN+; 0.8\% \textsf{CR} \\
\textsf{Student} &
48.7\% No; 2.5\% Yes; 48.7\% \textsf{Exp.} &
59.2\% No; 1.7\% Yes; 38.3\% \textsf{Exp.}; 0.8\% \textsf{CR} \\
\textsf{Employment} &
32.8\% Full-time; 5.9\% Part-time; 2.5\% Unemployed; 7.6\% No paid work; Others 0.8\%; 49.6\% \textsf{Exp.} &
32.5\% Full-time; 9.2\% Part-time; 8.3\% Unemployed; 9.2\% No paid work; 40.0\% \textsf{Exp.}; 0.8\% \textsf{CR} \\
\bottomrule
\end{tabularx}
\end{table}

We first report participation and abandonment statistics for each task. In the AV-Deepfake1M-based task, 130 workers were assigned a work unit. Of these, 120 completed it (92.31\%), while 10 did not complete within the time available (7.69\%), which we treat as abandonment following \citet{8873609}. In the TMC-based task, 144 workers were assigned a work unit; 120 completed it (83.33\%), while 24 did not complete (16.67\%). Two additional workers attempted to join the study but were not assigned a new work unit because they had already completed one. Overall, across both tasks, 240 out of 274 assigned workers completed their task (87.59\%), while 34 did not complete (12.41\%).

We report demographic statistics for the workers who completed the two tasks. Table~\ref{tab:demographics-v1-v2} summarizes both the questionnaire-based demographics (top section) and the Prolific-provided demographics (bottom section). Across both tasks, the sample skews toward mid-to-older adults. In the AV-Deepfake1M-based task, 43.3\% of workers are aged 36--50 and 22.5\% are aged 51--80, while in the TMC-based task these two ranges account for 38.3\% and 35.0\%, respectively. Education is medium-to-high in both tasks: the most common level is a bachelor's degree (36.7\% and 38.3\%), and a smaller but non-trivial share reports postgraduate or professional education (20.0\% and 15.8\%). Political affiliation and ideology are mixed, with the TMC-based task showing a higher share of Democrats (40.0\% vs.\ 33.3\%) and of workers identifying as liberal (32.5\% vs.\ 25.8\%). Reported income is mostly in the middle-to-upper brackets, with \$30k--\$75k being the most frequent bin in both tasks (45.0\% and 37.5\%).

The demographics provided by Prolific confirm that workers are almost entirely US-based and are mostly native English speakers. Almost all workers reside in the United States (100\% and 99.2\%), and most report English as their first language (89.1\% and 92.5\%). Gender is close to balanced (female: 52.1\% and 49.2\%; male: 47.9\% and 49.2\%). Ethnicity is mostly white, with a higher share of non-white categories in the TMC-based task than in the AV-Deepfake1M-based task (white: 67.5\% vs.\ 75.6\%). Finally, student and employment status include substantial missing values (mostly \textsf{Exp.}), so we interpret these fields descriptively.

Overall, the two tasks recruited broadly comparable samples on key factors, and the worker pool remained largely US-based. The abandonment rates align with those reported in prior crowdsourcing studies, and the demographic distributions are consistent with earlier work~\cite{roitero2020crowd,SOPRANO2021102710,10.1145/3531146.3534629,BARBERA2024103792,roitero2021crowd,10.1145/3340531.3412048,10.1145/3726302.3730091,10.1007/978-3-030-45442-5_26}.

\subsection{\ref{rq-1}: Accuracy of Authenticity Detection}
\label{sec:results-subsec:df-accuracy}

\begin{figure}[t]
  \centering

  % Left column: authenticity (stacked)
  \begin{minipage}[t]{0.3\linewidth}
    \centering
    \includegraphics[width=\linewidth]{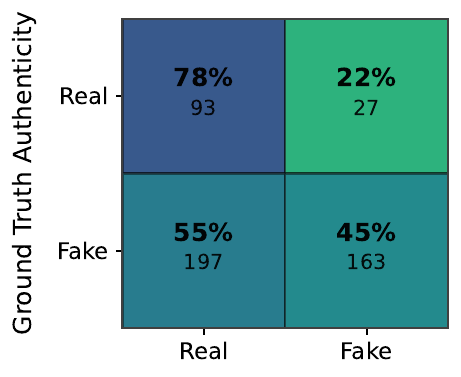}
    \phantomcaption\label{fig:heatmap-authenticity-v1}

    \includegraphics[width=\linewidth]{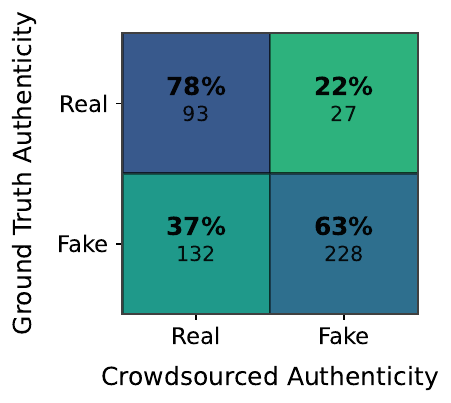}
    \phantomcaption\label{fig:heatmap-authenticity-v2}
  \end{minipage}
  % Right column: Manipulation (stacked)
  \begin{minipage}[t]{0.66\linewidth}
    \centering
    \includegraphics[width=.8\linewidth]{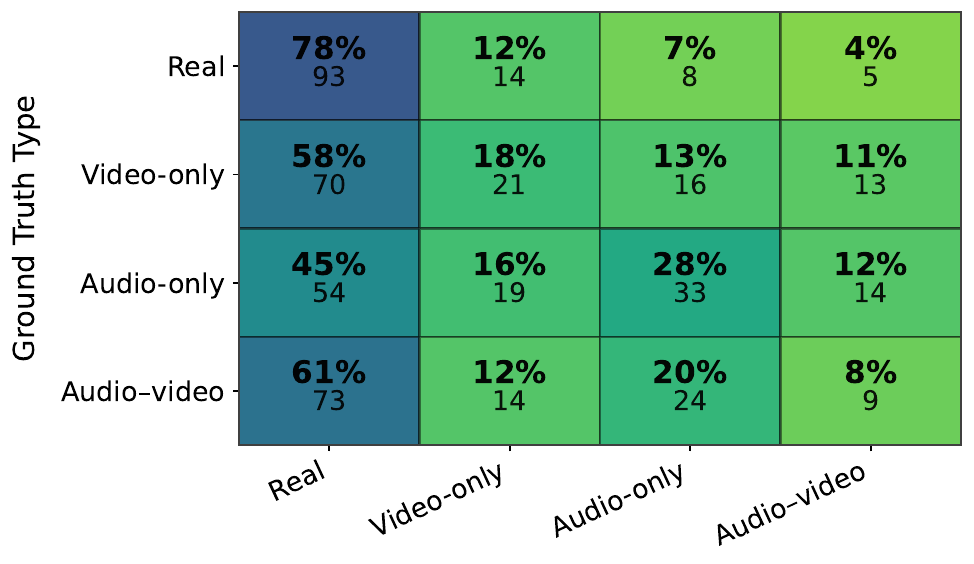}
    \phantomcaption\label{fig:heatmap-manipulation-v1}

    \includegraphics[width=.8\linewidth]{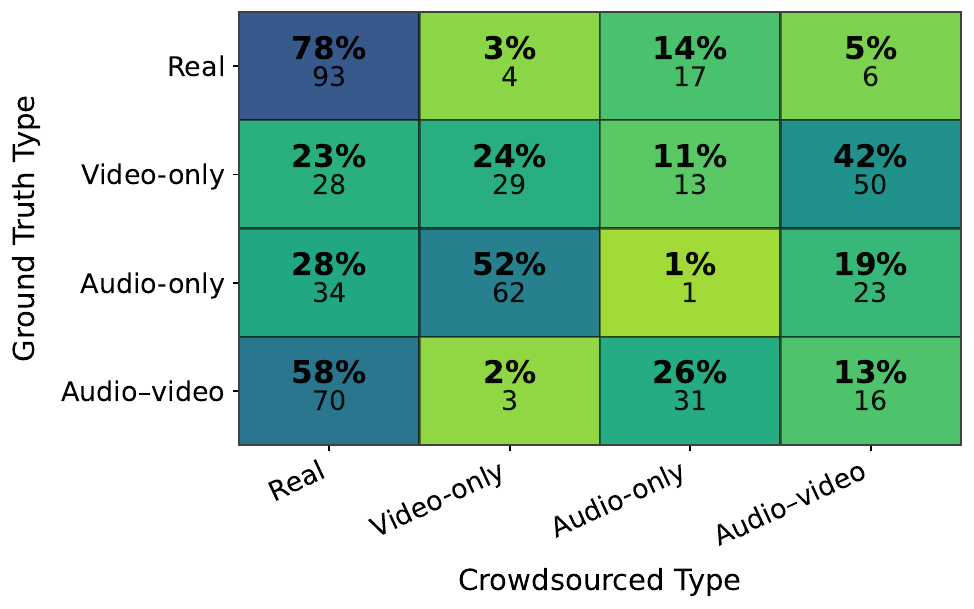}
    \phantomcaption\label{fig:heatmap-manipulation-v2}
  \end{minipage}

\caption{Individual worker judgments for AV-Deepfake1M (top) and TMC (bottom). Left: authenticity judgments. Right: manipulation type identification. Each cell reports the percentage within its ground truth row and the corresponding number of judgments. Darker cells indicate higher percentages.}  
\label{fig:heatmaps-all}
\end{figure}

Figure~\ref{fig:heatmaps-all} (left) shows individual authenticity judgments for AV-Deepfake1M (top) and TMC (bottom). Each heatmap summarizes the 480 judgments collected for the corresponding dataset (48 videos $\times$ 10 judgments). In the heatmaps, the authentic option is labeled \textsf{Real}; therefore, for ground truth manipulated videos, entries in the \textsf{Real} column correspond to missed manipulations, since type labels are provided only when workers judge a video as manipulated. In both datasets, errors are dominated by missed manipulations, especially on AV-Deepfake1M, where $54.7\%$ of judgments on manipulated videos are \textsf{Real}, compared to $36.7\%$ in TMC. Conversely, false positives on authentic videos are lower and identical across tasks ($22.5\%$).

Table~\ref{tab:authenticity-metrics-compact} summarizes the corresponding video-level performance after aggregation. Under majority vote, false positives remain low in both datasets (FPR $=0.083$), while false negatives are substantially higher and markedly worse on AV-Deepfake1M (FNR $=0.722$) than on TMC (FNR $=0.444$), confirming that missed manipulations are the primary driver of reduced performance, especially on AV-Deepfake1M.

\begin{table}[t]
\centering
\caption{Video-level authenticity detection metrics. For each dataset, a sample of 48 videos is used (12 authentic, 36 manipulated). Precision (P), recall (R), and F1 are reported for authentic and manipulated classes, together with overall accuracy (Acc.) and the corresponding error rates (FPR, FNR). Bold values indicate the best result within each dataset for Acc. (higher is better) and for FPR/FNR (lower is better).}
\label{tab:authenticity-metrics-compact}
\setlength{\tabcolsep}{4.2pt}
\begin{tabular}{@{}llccc ccc ccc@{}}
\toprule
\multirow{2}{*}{\textbf{Dataset}} &
\multirow{2}{*}{\textbf{Method}} &
\multicolumn{3}{c}{\textbf{Authentic}} &
\multicolumn{3}{c}{\textbf{Manipulated}} &
\multicolumn{3}{c}{\textbf{Overall}} \\
\cmidrule(lr){3-5}\cmidrule(lr){6-8}\cmidrule(lr){9-11}
& & \textbf{P} & \textbf{R} & \textbf{F1}
& \textbf{P} & \textbf{R} & \textbf{F1}
& \textbf{Acc.} & \textbf{FPR} & \textbf{FNR} \\
\midrule

\multirow{2}{*}{\textit{AV-Deepfake1M}}
& Majority Vote    & 0.297 & 0.917 & 0.449 & 0.909 & 0.278 & 0.426 & 0.438 & \textbf{0.083} & 0.722 \\
& Dempster-Shafer  & 0.281 & 0.750 & 0.409 & 0.812 & 0.361 & 0.500 & \textbf{0.458} & 0.250 & \textbf{0.639} \\
\addlinespace[3pt]
\cdashline{1-11}
\addlinespace[3pt]
\multirow{2}{*}{\textit{TMC}}
& Majority Vote    & 0.407 & 0.917 & 0.564 & 0.952 & 0.556 & 0.702 & 0.646 & \textbf{0.083} & 0.444 \\
& Dempster-Shafer  & 0.455 & 0.833 & 0.588 & 0.923 & 0.667 & 0.774 & \textbf{0.708} & 0.167 & \textbf{0.333} \\
\bottomrule
\end{tabular}
\end{table}

This missed-manipulation pattern also varies by manipulation type. In AV-Deepfake1M, manipulated videos are labeled as authentic in $45.0\%$ of audio-only judgments, $58.3\%$ of video-only judgments, and $60.8\%$ of audio-video judgments. In TMC, the corresponding rates are $28.3\%$, $23.3\%$, and $58.3\%$. This suggests that audio-video manipulations remain hard in both datasets, while single-modality cases are missed less often in TMC. While individual workers sometimes over-flag authentic videos, labeling $22.5\%$ of judgments on authentic videos as manipulated, aggregation reduces these errors at the video level because each authentic video receives ten judgments and most still have an authentic majority. Because each pool includes 12 authentic and 36 manipulated videos, overall accuracy reflects this class balance, effectively weighting authentic accuracy by $25\%$ and manipulated accuracy by $75\%$.

Under majority vote, TMC is substantially easier than AV-Deepfake1M. For manipulated videos, recall increases from $0.278$ to $0.556$ and F1 from $0.426$ to $0.702$. In contrast, recall on authentic videos is high and identical in both tasks ($0.917$), suggesting that the dataset difference is driven mainly by missed manipulations rather than by systematically over-flagging authentic content. This difference is also reflected in overall accuracy, which rises from $0.438$ (AV-Deepfake1M) to $0.646$ (TMC).

DS aggregation shifts the trade-off toward detecting more manipulations. In both datasets, recall on manipulated videos increases and manipulated F1 improves accordingly. In AV-Deepfake1M, manipulated recall rises from $0.278$ to $0.361$, while in TMC it rises from $0.556$ to $0.667$. However, this comes with more false alarms on authentic videos, reflected by a drop in authentic recall from $0.917$ to $0.750$ in AV-Deepfake1M and from $0.917$ to $0.833$ in TMC. Overall accuracy increases under DS in both datasets, consistent with the larger gains on manipulated videos under the 12/36 class balance. Following the procedure described in Section~\ref{sec:methodology-subsec:stats}, an exact McNemar test indicates that the difference between DS and majority vote is not significant in either dataset after Bonferroni correction ($p \ge 0.05$). Accuracy is nonetheless slightly higher under DS in both datasets: in AV-Deepfake1M it increases from $0.438$ under majority vote to $0.458$ under DS, and in TMC it increases from $0.646$ to $0.708$.

We also compare the per-video worker accuracy distributions between AV-Deepfake1M and TMC. Accuracy is higher on TMC (median $=0.800$) than on AV-Deepfake1M (median $=0.500$), and the difference is statistically significant under a two-sided Mann-Whitney U test ($p < 0.05$), with a small-to-moderate effect size (Cliff's $\delta=-0.296$)~\cite{mann1947test,cliff1993dominance}.

\subsection{\ref{rq-2}: Consistency of Authenticity Judgments}
\label{sec:results-subsec:consistency-crowd}

\begin{figure}[t]
  \centering
  \includegraphics[width=\linewidth]{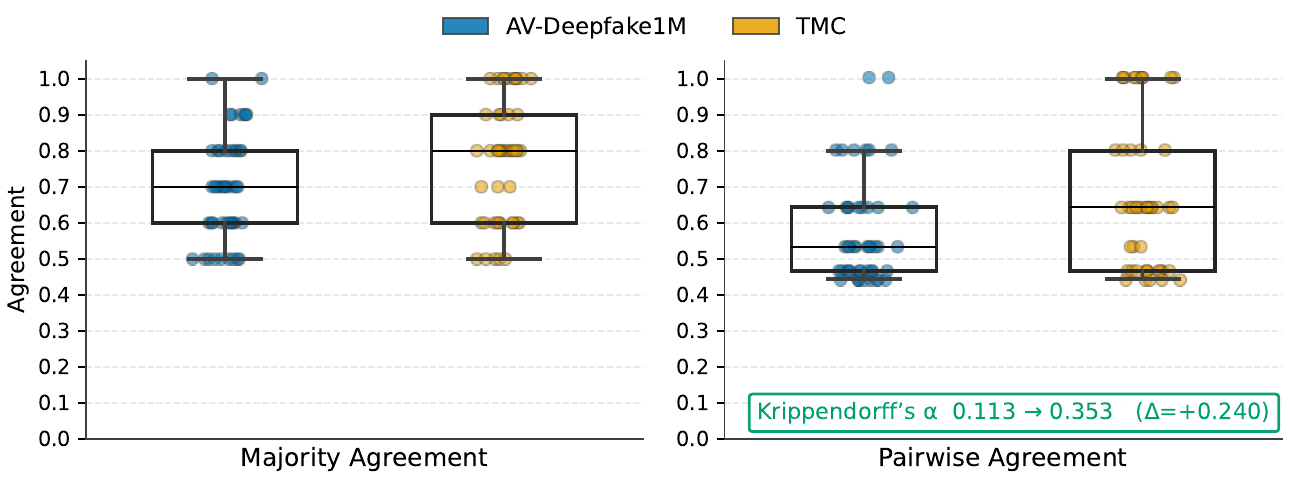}
\caption{Internal agreement of workers' authenticity judgments. Left: majority agreement. Right: pairwise agreement. Boxplots summarize per-video agreement scores across 48 videos per task, with each point corresponding to one video. Krippendorff's $\alpha$ is reported as an overall summary for each task.}
\label{fig:agreement-authenticity}
\end{figure}

Figure~\ref{fig:agreement-authenticity} summarizes the internal consistency of workers' authenticity judgments using the agreement measures introduced in Section~\ref{sec:metrics-subsec:agreement}. Krippendorff's $\alpha$ provides an overall summary for each task. The boxplots show per-video distributions of majority and pairwise agreement, with each point corresponding to one video.

Agreement is limited in both tasks, but it is consistently higher on TMC. In the AV-Deepfake1M-based task, we observe $\alpha=0.113$, mean pairwise agreement $=0.575$, and mean majority agreement $=0.696$, which implies that the modal label receives about $7/10$ votes per video on average. In the TMC-based task, all three measures increase to $\alpha=0.353$, mean pairwise agreement $=0.677$, and mean majority agreement $=0.777$, corresponding to about $8/10$ votes for the modal label. While $\alpha$ remains far from perfect agreement, the higher values on TMC indicate a more stable authenticity signal than on AV-Deepfake1M. Since pairwise agreement averages over all worker pairs, it is more sensitive to split votes and therefore decreases faster when judgments are more dispersed. In line with this, the smaller difference between mean majority and mean pairwise agreement in TMC ($0.100$) compared to AV-Deepfake1M ($0.121$) suggests that votes are more concentrated around the majority label in TMC.

We then compare the per-video agreement distributions between AV-Deepfake1M and TMC. For majority agreement, the median increases from $0.700$ to $0.800$; for pairwise agreement, it increases from $0.533$ to $0.644$. Both differences are significant under two-sided Mann-Whitney U tests after Bonferroni correction ($p<0.05$), with a small-to-moderate effect size (Cliff's $\delta=-0.279$), indicating higher agreement on TMC.

\subsection{\ref{rq-3}: Identification of the Manipulation Type}
\label{sec:results-subsec:manipulation-type}

Figure~\ref{fig:heatmaps-all} (right) summarizes individual judgments for manipulation type identification, with AV-Deepfake1M shown on top and TMC at the bottom; in both matrices, each row sums to $100\%$. Rows correspond to the ground truth condition, and columns to the four outcomes defined by the interface. The first column corresponds to an authentic judgment (\textsf{Real} in the interface); therefore, for ground truth manipulated videos, entries in that column represent missed manipulations.

Overall, manipulation typing is noisy in both datasets, largely because many manipulated videos remain labeled as \textsf{Real}. When considering all responses in the two-step interface and treating missed manipulations as \textsf{Real}, judgment-level accuracy on the four-label outcome space is low: micro-accuracy is $0.325$ for AV-Deepfake1M and $0.290$ for TMC. This is especially evident for audio-video cases, where $60.8\%$ of AV-Deepfake1M judgments and $58.3\%$ of TMC judgments are \textsf{Real}.

To better isolate the typing step from authenticity failures, we also analyze only judgments that include a type label, that is, cases where workers label a video as manipulated. Conditioning on these cases, AV-Deepfake1M shows reasonable attribution for single-modality manipulations (audio-only: $50.0\%$; video-only: $42.0\%$), whereas audio-video manipulations are rarely identified as such ($19.1\%$) and are most often reduced to a single modality, especially audio-only ($51.1\%$). In TMC, the same reduction pattern is even stronger and often shifts across modalities: audio-only is almost never labeled as audio-only ($1.2\%$) and is primarily labeled as video-only ($72.1\%$), while audio-video is most often reduced to audio-only ($62.0\%$). Overall, these patterns suggest that workers may notice that something is off, but often fail to attribute the anomaly to the affected channel.

\begin{figure}[t]
  \centering
  \includegraphics[width=\linewidth]{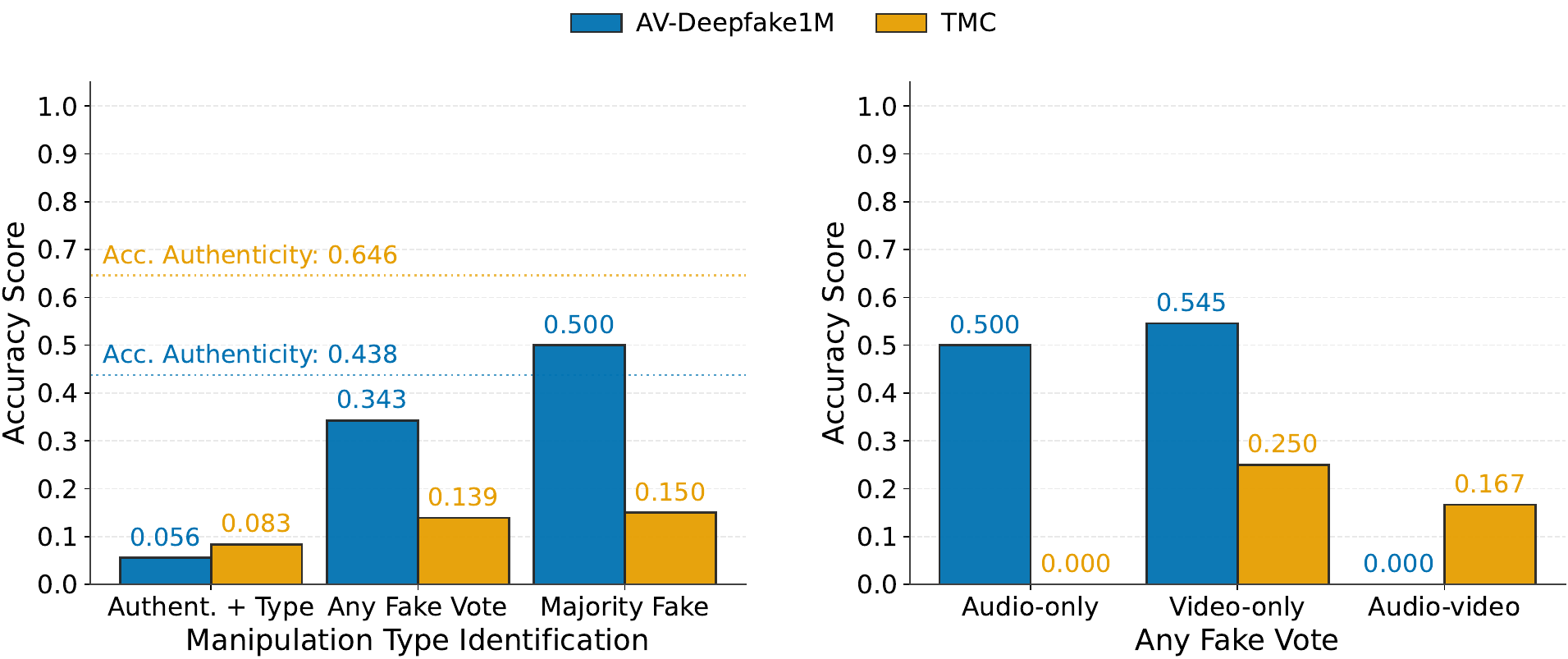}
    \caption{Video-level manipulation type identification accuracy on ground truth manipulated videos. Left: overall accuracy across evaluation settings. Right: accuracy by manipulation type under the \textsf{Any Fake Vote} setting. Dashed lines indicate overall authenticity accuracy (Table~\ref{tab:authenticity-metrics-compact}).}
  \label{fig:manipulation-type-accuracy}
\end{figure}

Figure~\ref{fig:manipulation-type-accuracy} reports video-level manipulation type identification accuracy on ground truth manipulated videos under the three evaluation settings described in Section~\ref{sec:metrics-subsec:classification}. In the left panel, \textsf{Authenticity + Type} leads to very low type accuracy for both datasets ($0.056$ for AV-Deepfake1M and $0.083$ for TMC), confirming that errors at the authenticity step propagate directly to typing. Under \textsf{Any Fake Vote}, coverage is almost complete ($0.972$ for AV-Deepfake1M and $1.000$ for TMC), but accuracy remains limited ($0.343$ and $0.139$), indicating that workers often disagree on whether the manipulation affects audio, video, or both even when at least one of them flags a manipulation. Restricting the evaluation to videos whose majority authenticity label is manipulated (\textsf{Majority Fake}) improves accuracy on AV-Deepfake1M ($0.500$) but leads to only a marginal change on TMC ($0.150$).

The right panel helps contextualize these outcomes. Under \textsf{Any Fake Vote}, AV-Deepfake1M shows higher accuracy for single-modality cases (audio-only: $0.500$; video-only: $0.545$) but collapses for audio-video ($0.000$), whereas TMC remains low across all modalities (audio-only: $0.000$; video-only: $0.250$; audio-video: $0.167$). Vote-level patterns can differ from the aggregated outcome because aggregation selects a single type even when votes are split across types.

We test whether typing difficulty differs across manipulation types within each dataset by treating each manipulated video as one observation and using the per-video typing accuracy rate under \textsf{Any Fake Vote}. In AV-Deepfake1M, the Kruskal-Wallis test indicates that typing accuracy differs across the three types ($p<0.01$). Post-hoc pairwise comparisons using Mann-Whitney tests with Bonferroni correction show that audio-video is harder than audio-only (median $=0.225$ vs.\ $0.450$; $p<0.05$), while the remaining pairs are not significant after correction. In TMC, the Kruskal-Wallis test again indicates differences across types ($p<0.001$), driven by audio-only being substantially harder than both video-only (median $=0.000$ vs.\ $0.250$; $p<0.001$) and audio-video (median $=0.000$ vs.\ $0.354$; $p<0.01$), while video-only and audio-video do not differ significantly after correction.

We also analyze the manipulation timestamps provided in manipulated judgments as a complementary cue when workers do flag a manipulation. On AV-Deepfake1M, the median IQR is $0.193$ and the median agreement is $0.333$, with $12.9\%$ of videos reaching agreement $\ge 0.6$ ($4/31$ videos). On TMC, timestamps are more concentrated, with a median IQR of $0.063$, a median agreement of $0.750$, and $75.8\%$ of videos reaching agreement $\ge 0.6$ ($25/33$ videos). A Mann-Whitney U test indicates a dataset effect for both IQR and agreement after Bonferroni correction ($p<0.01$). By manipulation type, agreement in TMC remains high for audio-only and video-only cases (median agreement $=0.800$ and $=0.764$) and decreases for audio-video ($=0.667$). In AV-Deepfake1M, agreement is lowest for audio-only (median agreement $=0.100$) and differs across types (Kruskal-Wallis $p<0.01$); post-hoc Mann-Whitney U tests with Holm correction show lower agreement for audio-only than for both video-only and audio-video ($p<0.01$).

\section{Discussion}
\label{sec:discussion}

We discuss the main findings by research question (\ref{rq-1}--\ref{rq-3}). Regarding \ref{rq-1}, crowdsourcing provides an authenticity signal for distinguishing authentic from manipulated videos, but many manipulations are still missed and performance depends strongly on the dataset (Figure~\ref{fig:heatmaps-all}, Table~\ref{tab:authenticity-metrics-compact}). Under the same protocol, TMC shows higher authenticity accuracy and stronger agreement than AV-Deepfake1M, suggesting that benchmark characteristics affect both correctness and the stability of the crowd signal. In practice, crowd judgments can support downstream screening and verification workflows, but their reliability should be validated on content that matches the target setting rather than assumed to generalize across benchmarks.

Regarding \ref{rq-2}, internal consistency on authenticity is limited overall, but it is higher on TMC than on AV-Deepfake1M (Figure~\ref{fig:agreement-authenticity}). This implies that dataset characteristics affect not only accuracy but also how concentrated votes are around the majority label, with implications for aggregation and for the reliability of crowd-based screening under dataset shift. In practice, agreement can serve as a lightweight uncertainty cue to prioritize low-consensus videos for further review.

As for \ref{rq-3}, manipulation typing remains the hardest part of the task (Section~\ref{sec:results-subsec:manipulation-type}, Figure~\ref{fig:manipulation-type-accuracy}). Even when workers judge a video as manipulated, they often struggle to attribute the anomaly to audio, video, or both, and joint audio-video cases are the most difficult to type. When cues are weak or spread across modalities, workers may rely on the most salient signal and reduce joint cases to a single-modality label. This points to a limitation of the current two-step interface: it supports the authenticity decision better than fine-grained modality attribution. At the same time, when workers do flag a manipulation, their timestamp reports can still converge on a plausible segment, which can help prioritize and focus downstream review even when modality attribution remains noisy. 

\section{Limitations}
\label{sec:limitations}

Several limitations should be considered when interpreting these results. Our evaluation is based on two balanced pools (48 videos per dataset) and ten judgments per video. This enables controlled comparisons, but the limited number of videos reduces statistical power and may not capture the full diversity of online videos encountered in practice.

Our sampling imposes feasibility constraints that can affect generalizability. We restrict candidate pools to videos that are playable in a browser-based interface and include both audio and video, and we stratify by manipulation condition to obtain balanced pools. While this improves comparability across datasets, it does not reflect real-world prevalence and may exclude content that is harder to render or inspect in crowdsourcing settings. Our participant pool is Prolific-based and largely US-based, and the task involves short, self-contained judgments under time and attention constraints. Results may differ with other populations, languages, or usage contexts (e.g., social media consumption, repeated exposure, or motivated reasoning). We also did not measure prior familiarity with deepfakes or related media literacy factors, which may contribute to individual variability.

Task onboarding may also have affected performance. While videos were audiovisual and the questionnaire made the multimodal nature explicit, the pre-task instructions did not explicitly prompt workers to keep audio enabled before the first item. Some workers may therefore have started with muted audio, potentially reducing sensitivity to audio-only and audio-video manipulations, especially early in the assignment. Aggregation can help when errors are not fully correlated across workers, but it cannot recover manipulations that most workers miss in the same way. Manipulation typing remains unreliable, particularly for joint audio-video cases, and the interface offers limited support for modality attribution beyond the type label.

Finally, we do not benchmark state-of-the-art automated deepfake detectors on the sampled pools. As a result, we cannot directly contextualize the observed dataset effects against current detector performance, nor use detector confidence or error profiles as an external proxy for sample hardness.

\section{Practical Implications}
\label{sec:implications}

From an operational perspective, these results suggest that crowdsourcing can provide an authenticity signal, with errors dominated by missed manipulations rather than false alarms. The choice of aggregation strategy induces a practical trade-off. Majority vote is conservative and keeps false alarms low, which is desirable when the cost of incorrectly flagging authentic content is high. Dempster-Shafer aggregation increases sensitivity to manipulations, but it also introduces additional false positives, which can be acceptable in recall-oriented screening settings where the goal is to surface candidates for downstream verification. By contrast, manipulation typing is currently too unstable to support downstream decisions on its own, especially for joint audio-video deepfakes (Section~\ref{sec:discussion}). Using crowd workers for modality attribution may require additional support, such as interfaces that make it easier to inspect both channels and focus on relevant segments.

These findings motivate a two-stage workflow. In the first stage, aggregated crowd judgments are used to screen videos as authentic or manipulated and prioritize items for further review. In the second stage, modality attribution and final verification are deferred to expert review or to model-assisted interfaces. Overall, crowd judgments appear most useful as a scalable screening signal that can be combined with downstream verification steps under dataset shift.

\section{Conclusions and Future Work}
\label{sec:conclusions-future}

These results support crowdsourcing as a scalable screening component for audiovisual deepfake detection when the task is well structured and highlight manipulation typing as the main barrier to reliable modality attribution. Across two matched tasks, we observe strong dataset effects in both accuracy and agreement, suggesting that the strength and stability of the crowd signal depend on dataset characteristics.

Future work will extend this study in several directions. We will test lightweight worker support, such as guided examples and short feedback, and evaluate whether it improves agreement~\cite{Diel2024HumanDeepfakeMetaAnalysis,somoray2023strategies}. We will also complement human judgments with automated baselines on the sampled pools to contextualize sample hardness and study when humans and detectors fail in similar or complementary ways.

We will build controlled datasets that vary modality, video duration, and manipulation strength to better understand which cues workers use and which cases are hardest to detect and type. We will extend the evaluation to multilingual and open-set audiovisual benchmarks to test robustness under language variation and broader manipulation conditions~\cite{hou2024polyglotfake,croitoru2025mavosdd}. We will use power-based sample size planning to estimate how many independent crowd judgments per video are needed to detect meaningful differences with a chosen statistical power~\cite{10.1145/3597201}. Finally, we will explore richer interfaces (e.g., keyframe highlighting, audio waveforms, and video timelines) and human-AI setups where models highlight uncertain segments to guide workers' attention.

\section*{Acknowledgments}

This research is partially supported by the PRIN 2022 Project -- ``MoT---The Measure of Truth: An Evaluation-Centered Machine-Human Hybrid Framework for Assessing Information Truthfulness'' -- Code No. 20227F2ZN3, CUP No. G53D23002800006, funded by the European Union -- Next Generation EU -- PNRR M4 C2 I1.1. We thank Kevin Roitero for feedback on misinformation and crowdsourcing, and Thomas Trigatti for his thesis work, which contributed to the development of this line of research.

\bibliographystyle{unsrtnat}
\bibliography{0-bibliography}

\clearpage

\appendix

\end{document}